\documentclass[a4paper]{article}
\usepackage{tikz}
\usetikzlibrary{matrix}
\usepackage{amscd,amsmath,amssymb}
\usepackage{bbm}

\newcommand{\so}{\mathfrak{so}}
\newcommand{\g}{\mathfrak{g}}
\newcommand{\su}{\mathfrak{su}}
\newcommand{\G}{\mathfrak{G}}
\newcommand{\mF}{\mathfrak F}

\newcommand{\p}[1]{(\ref{#1})}

\newcommand{\cW}{{\cal W}}

\newcommand{\hJ}{{\hat J}}

\newcommand{\eps}{\varepsilon}

\newcommand{\cN}{{ {\cal N}   }}

\newcommand{\tW}{{\widetilde W}}
\newcommand{\tV}{{\widetilde V}}

\def\hX{\widehat{X}}
\def\hZ{\widehat{Z}}

\def\hW{\widehat{W}}
\def\hY{\widehat{Y}}
\def\hJ{\widehat{J}}
\def\hV{\widehat{V}}

\newcommand{\be}{\begin{equation}}
\newcommand{\ee}{\end{equation}}
\newcommand{\bea}{\begin{eqnarray}}
\newcommand{\eea}{\end{eqnarray}}

\newcommand{\ba}{\begin{aligned}} \newcommand{\ea}{\end{aligned}}

\def\im{{\rm i}}

\newcommand{\nn}{\nonumber}
\def\theequation{\arabic{section}.\arabic{equation}}

\begin{document}
\begin{flushright}
%\today\\
%Version 5.0 -semi-final release
\end{flushright}\vspace{1cm}
\begin{center}
{\Large\bf Two faces of  $\cN=7,8$  superconformal  mechanics}
\end{center}
\vspace{1cm}

\begin{center}
{\large\bf  Sergey Krivonos${}^{a,b}$ and Armen Nersessian${}^{a,c,d}$}
\end{center}

\vspace{0.2cm}

\begin{center}
{${}^a$ \it
Bogoliubov  Laboratory of Theoretical Physics, JINR,
141980 Dubna, Russia}

${}^b${\it Laboratory of Applied Mathematics and Theoretical Physics, TUSUR, Lenin. Ave 40, 634050 Tomsk, Russia}

${}^c$ {\it Yerevan Physics Institute,
	2 Alikhanyan Brothers  St., Yerevan, 0036, Armenia}

${}^d$ {\it Institute of Radiophysics and Electronics, Ashtarak-2, 0203, Armenia }
\vspace{0.5cm}

{\tt krivonos@theor.jinr.ru}, {\tt arnerses@yerphi.am}
\end{center}
\vspace{2cm}

\begin{abstract}\noindent
Two variants of $\cN=7$ superconformal mechanics with manifest $\so(7)$ and $\su(2) \times \su(2)$ $R$-symmetries possessing  exceptional $G(3)$ dynamical symmetry are presented.
To construct these $G(3)$ theories, we developed two new versions of covariant embedding
of the exceptional $\g_2$ algebra into the $\so(7)$ algebra and then the embedding of $\so(7)$ into $\so(8)$. As a result, two $\cN=8$ superconformal mechanics with the $OSp(8|2)$ and $\mF(4)$ superconformal algebras and $\cN=7, G(3)$ superconformal mechanics were constructed in a uniform way. The constants of the octonion multiplication play a key role in the construction of superconformal mechanics with manifest $\so(7)$ symmetry.
\end{abstract}

\vskip 1cm
\noindent
%PACS numbers: 11.30.Pb, 11.30.-j

\vskip 0.5cm

\noindent
%Keywords: Calogero models , $N$--extended supersymmetry, Integrable system, super-integrability

\newpage

\setcounter{equation}{0}
\section{Introduction}

Superconformal mechanics with an $\so(1,2)$ conformal symmetry  are potentially of great interest due to their possible relation to AdS${}_2$ holography. In virtue of the AdS/CFT correspondence, there is by now a large amount of works on anti-de Sitter solutions of string/M-theory in diverse dimensions.  The high dimensionality of the internal space of
AdS${}_2$ solutions of string/M-theory is related via duality with $R$-symmetry
of the corresponding superconformal mechanics. However, all variants of $\cN=4$ superconformal mechanics possess the exceptional superalgebra $D(1,2;\alpha)$ as  dynamical symmetry. Therefore, the maximal $R$-symmetry for $\cN=4$ cases is $\so(1,2) \times \so(4)$. 

Passing to  $\cN=8$ superconformal mechanics makes the situation more interesting because
in contrast with the $\cN=4$ case in the $\cN=8$ supersymmetric case we have four different superconformal algebras $OSp(8|2), \, \mF(4), \, OSp(4^\star |4)$, and $SU(1,1|4)$ (see e.g. \cite{VP}).  Thus, the $R$-symmetries are extended to $\so(8), \so(7), \so(5)\times \su(2)$ and $\su(4)\times {\mathfrak u}(1)$, respectively. Many variants  of $\cN=8$ superconformal mechanics were constructed, mainly in the superfield Lagrangian formalism [2-9] and within the Hamiltonian  approach as well [10-16]. In the case of $\cN=8$ superconformal models,    the Hamiltonian approach  has several preferences, such as
\begin{itemize}
	\item In all variants of  superconformal mechanics with a different number of supersymmetries 
	the superconformal symmetry is spontaneously broken, as the result of which the dilaton $u$ plays a master role in the form of supercharges and Hamiltonian. Roughly speaking, the supercharges have the form  (cf. \cite{hkn1,KN1,SQS24,anton,tigran})
	\be\label{01} Q \sim p_r \psi + \frac{1}{r} \left[ R-symmetry \; generators\right] \psi .\ee
	Here $r=e^{u/2}$ and $p_r$ is the corresponding momentum.
	\item The explanation of this form \p{01} is quite simple: due to the presence of the dilaton 	fields, one can always realize the dilaton ${\cal D}$, conformal boost ${\cal K}$, and the conformal supersymmetry ${\cal S}$  generators as
	\be\label{02}
	{\cal D}= \frac{1}{2} r p_r, \quad {\cal K}= \frac{1}{2}r^2, \qquad {\cal S} =  r \psi .
	\ee
	\item Thus, the form of the supercharges \p{01} is unique  to produce the main relations of any superconformal algebra
	\be
	\{Q,{\cal S}\} \sim {\cal D} + \left[ R-symmetry \; generators\right].
	\ee
\end{itemize}
As a result, all peculiarities of different variants of superconformal mechanics are hidden in the realization of $R$-symmetry generators, mainly in the bosonic but partially in the fermionic realizations too. Specializing on the first two superconformal mechanics
with $OSp(8|2)$ and $\mF(4)$ superalgebras, we have to deal with (at least) two realizations
of $\so(8)$ and $\so(7)$ algebras  with manifest $\so(7)$ and $\su(2)^4 \mbox{ and } \su(2)^3$ symmetries. Moreover, this line can be continued as follows:
$$
OSp(8|2) \rightarrow \mF(4) \rightarrow G(3) \quad \leftrightarrow \quad
\so(8) \rightarrow \so(7) \rightarrow \g_2 .
$$
The last case in this series is the $\cN=7$ superconformal mechanics which is based, as it was stated in \cite{FT1}, on $(1,7,7,1)$ supermultiplet\footnote{In \cite{FT1},  the quantum version of  $G(3)$ superconformal mechanics was considered.}. In the present paper we provide a detailed uniform description of these superconformal mechanics using the properly defined line of embeddings $\so(8) \rightarrow \so(7) \rightarrow \g_2$.  The manifest $\so(7)$ covariant description developed in this paper reveals the tight relations of the $\cN=8$ superconformal algebras with the algebra of octonions. Indeed, the supercharges we constructed contain the constants of the octonion multiplication $c_{ijk}$. Of course, the relation of the exceptional algebras with the octonions hs been known for a long time\footnote{In the present context such relations were discussed in \cite{toppan3,FT1,SQS24}.}  However, explicit appearance of constants of the octonion multiplication $c_{ijk}$ and their dual cousins $f_{ijkl}$ \p{f} in the supercharges and Hamiltonian has not been noted yet, to the best of our knowledge, as well as explicit expressions of embedding of $\g_2$ into $\so(7)$ \p{G2}.

The plan of our paper is as follows. In  section 2 we describe two realizations of the algebra $\so(8)$ and its $\so(7)$ and $\g_2$ subalgebras paying much attention to the explicit embeddings $\g_2 \rightarrow \so(7) \rightarrow \so(8)$. In Section 3 we explicitly construct supercharges and  Hamiltonians for  $OSp(8|2), \, \mF(4), G(3)$
superconformal mechanics within the $\so(7)$ manifestly covariant formalism. In  Section 4 we repeat the construction of superconformal mechanics within the $\su(2)$ invariant description. In Conclusion we review the obtained results and mention possible future developments. In Appendix, we present the explicit variant of changing variables from the $\su(2)$ invariant formalism to the commonly adopted one.

\setcounter{equation}{0}
\section{Two faces of  $\so(8)$ and its subalgebras}

\subsection{Standard realization}
The standard realization of the algebra $\so(8)$ means its realization in the
sixteen-dimensional bosonic phase spaces, and eight-dimensions fermionic one. Thus, let  us introduce the  following set of bosonic and fermionic fields
\begin{itemize}
	\item sixteen bosonic fields:  $y_0, p_0$ and $y_i, p_j$ $i,j = 1,\ldots, 7$
	\item eight fermionic fields: $\psi_0, \psi_i$
\end{itemize}
which obey the following non zero Poisson brackets:
\be\label{PB}
 \big\{ p_0,y_0 \big\} =1\, ,\quad  \big\{ p_i,y_j \big\} = \delta_{ij}\,,\qquad\big\{ \psi_0, \psi_0 \big\} = \im\,,\quad\big\{ \psi_i, \psi_j \big\} = \im \; \delta_{ij}\; .
\ee

\subsubsection{$\so(7)$ and $\so(8)$ algebras}
From our fields one can construct currents spanning two different $\so(7)$ algebras:
bosonic and fermionic:
\bea\label{so7}
W_{ij}  =  p_i y_j -p_j y_i  &\rightarrow &\quad
\{W_{ij}, W_{kl}\}  = \delta_{ik} W_{jl}-\delta_{jk} W_{il}-\delta_{il}W_{jk}+\delta_{jl} W_{ik}, \nn \\
\hW_{ij} = \im\,\psi_i \psi_j &\rightarrow &\quad \{\hW_{ij}, \hW_{kl}\} = \delta_{ik} \hW_{jl}-\delta_{jk} \hW_{il}-\delta_{il} \hW_{jk}+\delta_{jl} \hW_{ik}.
\eea
To construct the generators of $\so(8)$ algebras one has to add to the generators $W_{ij}$ and $\hW_{ij}$ seven bosonic $V_i$ and seven fermionic generators $\hV_i$, respectively, transforming under 7-dimensional representations of $\so(7)$:
\bea\label{so8}
V_i = p_0 y_i-p_i y_0 \, &\rightarrow &\quad
\{W_{ij}, V_{k}\} = \delta_{ik} V_{j}-\delta_{jk} V_{i}, \quad 
\{V_{i}, V_{j}\} = W_{ij}, \nn \\
\hV_i = \im\,\psi_0 \psi_i \, &\rightarrow &\quad
\{\hW_{ij}, \hV_{k}\} = \delta_{ik} \hV_{j}-\delta_{jk} \hV_{i}, \quad 
\{\hV_{i}, \hV_{j}\} = \hW_{ij}.
\eea

\subsubsection{$\mathfrak{G}_2$ algebra}
The embedding of $\g_2$ algebra into $\so(7)$  can be achieved by the rank 2 totally antisymmetric tensor $\cW_{ij}$ constructed from the $\so(7)$ generators \p{so7} and obeying the constraints \cite{FT1}
\be\label{maing2}
c_{ijk} \cW_{jk}=0 \, .
\ee

Here the rank-3 totally antisymmetric tensor $c_{ijk}$ is the structure constant
of the  octonionic multiplication. Namely, if the $e_i, i=1,\ldots,7$ are seven octonions with the following multiplication:
\be
e_i e_j = -\delta_{ij} + c_{ijk} e_k ,
\ee
the tensor $c_{ijk}$ has the following non zero components\footnote{Here we followed the notation adopted in  \cite{FT1}.}
\be
c_{123}=c_{147}=c_{165}=c_{246}=c_{257}=c_{354}=c_{367}=1.
\ee
We will also need the dual, rank-4  totally antisymmetric tensor $f_{ijkl}$ defined as follows
\be\label{f}
f_{ijkl} =\frac{1}{6} \epsilon_{ijklmnp} c_{mnp}, \quad \epsilon_{1234567}=1.
\ee

It is rather easy to check that the solution of equations \p{maing2} $\cW_{ij}$ can be constructed as follows\cite{dewit}:
\be\label{G2}
\cW_{ij} = W_{ij} +\frac{1}{4} f_{ijkl} W_{kl} .
\ee
The tensor $\cW_{ij}$ obeys the following Poisson brackets:
\be
\{ \cW_{ij}, \cW_{kl}\} =-\frac{3}{2} \left( f_{ijkm} \cW_{lm} -f_{ijlm}\cW_{km} \right) +
c_{ijm}c_{kln} \cW_{mn} .
\ee
Thus, fourteen currents $\cW_{ij}$ span $\g_2$ algebra.

\subsubsection{Non-standard embedding of $\so(7)$ into $\so(8)$}

With the help of the constants $c_{ijk}$ and $f_{ijkl}$ one can define two different embeddings of $\so(7)$ into $\so(8)$:
\bea
&& \tW_{ij} =\frac{1}{2}\left(W_{ij} + \alpha c_{ijk} V_k+\frac{1}{2} f_{ijkl}W_{kl}\right), \quad \alpha^2=1  \label{tW}\\
&& \{\tW_{ij}, \tW_{kl}\} = \delta_{ik} \tW_{jl}-\delta_{jk} \tW_{il}-\delta_{il} \tW_{jk}+\delta_{jl} \tW_{ik}. \nn
\eea
The additional generator 
\be\label{tV}
\tV_i =\frac{1}{2} \left( V_i +\frac{\alpha}{2} c_{ijk} W_{jk}\right)
\ee
with the new defined generators $\tW_{ij}$ form the $\so(8)$ algebra
\be
\{\tW_{ij}, \tV_{k}\} = \delta_{ik} \tV_{j}-\delta_{jk} \tV_{i}, \qquad 
\{\tV_{i}, \tV_{j}\} = \tW_{ij}.
\ee

\subsection{Realization with manifest $\su(2)$ symmetries}
Such realization was used in the papers \cite{fed1,hkn1,KN1}. To construct the generators of the $\so(8)$ algebra, we again need sixteen bosonic and eight fermionic fields but now with different symmetry properties - all fields are supposed to transform under fundamental representations of different $\su(2)$ algebras. Thus, we will  consider the following set of bosonic and fermionic fields:
\begin{itemize}
	\item sixteen  bosonic fields: $B^{a\,\alpha} {}_m$ and $Z^{i\,A}{}_m,\quad \{a,\,\alpha,\, i, \,A, \,m\} = 1,2$
	\item  eight fermionic fields $\phi^{i\,A}, \chi^{a\,\alpha}$
\end{itemize}
such that satisfy the following conjugation rule:
\be\label{conj}
(Z^{i A}{}_m)^{\dagger} = Z_{i}{}_{A m}\,, \quad (B^{a \alpha}{}_m)^{\dagger} =B_{a\,\alpha\,m}\,, \qquad  (\phi^{i\,A})^{\dagger}  = \phi_{i\,A},\quad (\chi^{a\,\alpha})^{\dagger}  = \chi_{a\,\alpha}
\ee
and for which the Poisson brackets hold
\bea\label{PB}
&&\big\{ Z^{i\,A}{}_m, Z^{j\,B}{}_n \big\}= 2\,\eps^{ij} \eps^{AB} \eps_{mn}\,, \qquad
\big\{ \phi^{i\,A}, \phi^{j\,B} \big\} =  2\,\im\, \eps^{ij} \eps^{AB}\,. \nn \\
&&\big\{ B^{a\,\alpha}{}_m, B^{b\,\beta}{}_n \big\}= 2\, \eps^{ab} \eps^{\alpha \beta} \eps_{mn}\,, \qquad
\big\{ \chi^{a\,\alpha}, \chi^{b\,\beta} \big\} =2\, \im\, \eps^{ij} \eps^{AB}\,.
\eea
Note, the indices $m,n$ are not the indices of any $\su(2)$ algebras, instead these indices just count the fields involved. However, the corresponding metric is $\epsilon^{nm}$, i.e. $A^n B_n$ means $\epsilon^{nm} A_m B_n$.
 
\subsubsection{$\so(8)$ algebra}
From our fields one can construct bosonic and fermionic  currents spanning four  different $\su(2)$ algebras:\\
four bosonic 
\bea
J^{ij} =\frac{1}{4} Z^{i \alpha\, m}Z^j{}_{\alpha\,m} \quad &\rightarrow &\quad
\{J^{ij}, J^{kl}\} =-\epsilon^{ik} J^{jl}-\epsilon^{jl}J^{ik}, \nn \\
W^{AB} =\frac{1}{4} B^{a A\, m}B_a{}^B{}_m \quad &\rightarrow &\quad
\{W^{AB},W^{CD} \} =-\epsilon^{AC}W^{BD} -\epsilon^{BD} A^{AC}, \nn \\
X^{a b}= \frac{1}{4} B^{a A\,m}B^b{}_{A\,m} \quad &\rightarrow &\quad
\{X^{ab},X^{cd}\} = -\epsilon^{ac}X^{bd} -\epsilon^{bd} X^{ac}, \nn \\
Y^{\alpha\beta}= \frac{1}{4} Z^{i\alpha\,m}Z_i{}^\beta{}_m \quad &\rightarrow &\quad
\{Y^{\alpha\beta}, Y^{\gamma \delta}\} = -\epsilon^{\alpha\gamma} Y^{\beta \delta} - \epsilon^{\beta \delta} Y^{\alpha \gamma} , \label{su2bos}
\eea
and four fermionic ones
\bea\label{su2fer}
\hJ^{ij} = \frac{\im}{4}  \phi^{i A} \phi^j{}_A  &\quad \Rightarrow \quad & \left\{\hJ^{ij}, \hJ^{kl}\right\} = -\epsilon^{ik} \hJ^{jl} -\epsilon^{jl}\hJ^{ik} , \nn \\
\hW^{A B} = \frac{\im}{4} \phi^{i A} \phi_i{}^B &\quad \Rightarrow \quad & \left\{\hW^{A B}, \hW^{C D}\right\} = -\epsilon^{A C} \hW^{B D} -\epsilon^{B D}\hW^{A C} , \nn \\
\hX^{a b } = \frac{\im}{4} \chi^{a \alpha} \chi^{b}{}_\alpha & \Rightarrow & 
\left\{\hX^{a b}, \hX^{c d}\right\} = -\epsilon^{a c} \hX^{b d} -\epsilon^{b d}\hX^{a c}, \nn \\
\hY^{\alpha \beta}=\frac{\im}{4} \, \chi^{a \alpha} \chi_{a}{}^\beta  &\quad \Rightarrow \quad & 
\left\{\hY^{\alpha \beta}, \hY^{\gamma \delta}\right\} = -\epsilon^{\alpha \gamma} \hY^{\beta \delta} -\epsilon^{\beta \delta}\hY^{\alpha \gamma}.
\eea
Let us define  sixteen bosonic currents $V^{i\,A\,a\,\alpha}$ and sixteen fermionic currents $\hV^{i\,A\,a\,\alpha}$ 
\be\label{so8}
V^{i\,A\,a\,\alpha}= Z^{i\,\alpha\,m}B^{a\,A}{}_m, \qquad \hV^{i A a \alpha} = \im\, \phi^{i A} \chi^{a \alpha}.
\ee
These fields transform under four $\su(2)$ generators \p{su2bos}  as follows 
\bea
\{J^{ij},V^{k \,A \, a\, \alpha}\} & = & -\frac{1}{2} \left( \epsilon^{ik} V^{j \,A \, a\, \alpha} +\epsilon^{jk} V^{i \,A \, a\, \alpha}\right), \quad
\{W^{AB},V^{i \,C \, a\, \alpha}\}  =  -\frac{1}{2} \left( \epsilon^{AC} V^{i \,B \, a\, \alpha} +\epsilon^{BC} V^{i \,A \, a\, \alpha}\right),\nn \\
\{X^{ab},V^{i \,A \, c\, \alpha}\} & = & -\frac{1}{2} \left( \epsilon^{ac} V^{i \,A \, b\, \alpha} +\epsilon^{bc} V^{i \,A \, a\, \alpha}\right), \quad
\{Y^{\alpha\beta},V^{i \,A \, a\, \gamma}\}  =  -\frac{1}{2} \left( \epsilon^{\alpha\gamma} V^{i \,A \, a\, \beta} +\epsilon^{\beta\gamma} V^{i \,A \, a\, \alpha}\right) .
\eea
These sixteen generators $,V^{k \,A \, a\, \alpha}$  form the $\so(8)$ algebra together with four $\su(2)$ algebras \p{su2bos}. The main Poisson brackets read
\bea\label{VVso8}
\left\{V^{i A a \alpha},V^{j B b \beta}\right\} = -4 \left( \epsilon^{AB}\epsilon^{ab} \epsilon^{\alpha \beta} J^{ij} + \epsilon^{ij}\epsilon^{ab} \epsilon^{\alpha \beta} W^{AB}+ \epsilon^{ij}\epsilon^{AB} \epsilon^{\alpha \beta} X^{ab} + \epsilon^{ij}\epsilon^{AB} \epsilon^{a b} Y^{\alpha \beta}\right). 
\eea

The hatted pure fermionic generators obey the same Poisson brackets.
\subsubsection{$\so(7)$ algebra}
Now the situation with embedding of the algebra $\so(7)$ into the algebra $\so(8)$ is more subtle than it was in the previous case in subsection 2.1.1. To provide such embedding, one has 
\begin{itemize}
	\item to identify two $\su(2)$ subalgebras, for example 
	\be\label{zso7}
	W^{AB} \oplus X^{ab} \rightarrow Z^{ab}_{\so(7)}=W^{ab}+X^{ab}
	\ee
	\item to symmetrize the generators $V^{i\,A\,a\,\alpha}$ over the corresponding indices: 
	\be\label{vso7}
	V^{i\, a\, b\, \alpha}_{\so(7)} = V^{i\, a\, b\, \alpha}+V^{i\, b\, a\, \alpha}.
	\ee
\end{itemize}
Thus, we have nine generators in three $\su(2)$ subalgebras $J^{ij},Z^{ab}_{\so(7)},Y^{\alpha\beta}$ and twelve generators  $V^{i\, a\, b\, \alpha}_{\so(7)}$ which together give a full 21-dimensional set of $\so(7)$ generators.

The new Poisson brackets read
\bea\label{su2so7}
\{ Z^{ab}_{\so(7)}, V^{i\,c\,d\,\alpha}_{\so(7)} \} & = & - \epsilon^{ac} V^{i\,b\,d\,\alpha}_{\so(7)}- \epsilon^{bd} V^{i\,c\,a\,\alpha}_{\so(7)},  \\
\{V^{i\,a\,b\,\alpha}_{\so(7)}, V^{j\,c\,d\,\beta}_{\so(7)}\} &=& -8 \left[ 
\epsilon^{ij} \epsilon^{\alpha\beta} \left( \epsilon^{ac} Z^{bd}_{\so(7)} + \epsilon^{bd} Z^{ac}_{\so(7)}\right)+\left(\epsilon^{ac}\epsilon^{bd}+\epsilon^{ad}\epsilon^{bc}\right) \left(\epsilon^{ij}Y^{\alpha\beta}+\epsilon^{\alpha\beta}J^{ij}\right)\right].\nn
\eea

The hatted pure fermionic generators have the same structure and obey the same Poisson brackets.
\subsubsection{$\mathfrak{G}_2$ algebra} 
To extract the generators of the $\g_2$ algebra from the generators of $\so(7)$, one can do the following:
\begin{itemize}
	\item The rank of the $\g_2$ algebra is equal to 2. Thus, one has to pick up two $\su(2)$ subalgebras the from $\so(7)$ algebra.
	So let us identify the indices $a,b,\alpha$. Therefore, instead of three $\su(2)$ algebras with the generators $J^{ij},Z^{ab}_{\so(7)},Y^{\alpha\beta}$ we  have two $\su(2)$ algebras with the generators
	\be\label{G21}
	J^{ij} \quad \mbox{ and } \quad Z^{ab}_{\g_2} = Z^{ab}_{\so(7)}+Y^{ab} = X^{ab}+W^{ab}+Y^{ab}\ee
	\item In the generators $V_{\so(7)}^{i a b \alpha}$ one has to identify and then  symmetrize the indices $a, b, \alpha$
	\be\label{G22}
	V_{\g_2}^{i a b c} =V^{i a b c}+V^{i b c a}+V^{i c a b}+V^{i b a c}+V^{i a c b}+V^{i c b a}
	\ee
	\item The generators $J^{ij}, Z_{\g_2}^{ab}$ and $V_{\g_2}^{i a b c}$ form the $\g_2$ algebra.
\end{itemize}
The main Poisson brackets read
\bea\label{g2}
\{Z_{\g_2}^{ab},V_{\g_2}^{i c d e}\}&=& -\frac{1}{2}\left(\epsilon^{ac} V_{\g_2}^{i b d e}+\epsilon^{bc} V_{\g_2}^{i a d e}+\epsilon^{ad} V_{\g_2}^{i c b e}+\epsilon^{bd} V_{\g_2}^{i c a e} +\epsilon^{ae} V_{\g_2}^{i c d b}+\epsilon^{be} V_{\g_2}^{i c d a}\right), \nn \\
\{V_{\g_2}^{i a b c},V_{\g_2}^{j d e f} \}& = & -8 \epsilon^{ij} \left[ \left(\epsilon^{be}\epsilon^{cf}+\epsilon^{bf}\epsilon^{ce}\right) Z_{\g_2}^{ad} +
\left(\epsilon^{bd}\epsilon^{cf}+\epsilon^{bf}\epsilon^{cd}\right) Z_{\g_2}^{ae} + 
\left(\epsilon^{bd}\epsilon^{ce}+\epsilon^{cd}\epsilon^{be}\right) Z_{\g_2}^{af} + \right. \nn \\
&& \left(\epsilon^{ae}\epsilon^{cf}+\epsilon^{af}\epsilon^{ce}\right) Z_{\g_2}^{bd} +
\left(\epsilon^{ad}\epsilon^{cf}+\epsilon^{af}\epsilon^{cd}\right) Z_{\g_2}^{be} +
\left(\epsilon^{ad}\epsilon^{ce}+\epsilon^{cd}\epsilon^{ae}\right) Z_{\g_2}^{bf} + \nn \\
&& \left.  \left(\epsilon^{be}\epsilon^{af}+\epsilon^{bf}\epsilon^{ae}\right) Z_{\g_2}^{cd} +
\left(\epsilon^{bd}\epsilon^{af}+\epsilon^{bf}\epsilon^{ad}\right) Z_{\g_2}^{ce} +
\left(\epsilon^{bd}\epsilon^{ae}+\epsilon^{ad}\epsilon^{be}\right) Z_{\g_2}^{cf} \right]- \nn \\
&& 8 J^{ij}  \left[ \left(\epsilon^{be}\epsilon^{cf}+\epsilon^{bf}\epsilon^{ce}\right) \epsilon^{ad} +
\left(\epsilon^{bd}\epsilon^{cf}+\epsilon^{bf}\epsilon^{cd}\right) \epsilon^{ae} + 
\left(\epsilon^{bd}\epsilon^{ce}+\epsilon^{cd}\epsilon^{be}\right)\epsilon^{af} + \right. \nn \\
&& \left(\epsilon^{ae}\epsilon^{cf}+\epsilon^{af}\epsilon^{ce}\right) \epsilon^{bd} +
\left(\epsilon^{ad}\epsilon^{cf}+\epsilon^{af}\epsilon^{cd}\right) \epsilon^{be} +
\left(\epsilon^{ad}\epsilon^{ce}+\epsilon^{cd}\epsilon^{ae}\right) \epsilon^{bf} + \nn \\
&& \left.  \left(\epsilon^{be}\epsilon^{af}+\epsilon^{bf}\epsilon^{ae}\right) \epsilon^{cd} +
\left(\epsilon^{bd}\epsilon^{af}+\epsilon^{bf}\epsilon^{ad}\right) \epsilon^{ce} +
\left(\epsilon^{bd}\epsilon^{ae}+\epsilon^{ad}\epsilon^{be}\right) \epsilon^{cf} \right].
\eea

The hatted pure fermionic generators have the same structure and obey the same Poisson brackets.

\setcounter{equation}0
\section{Superconformal mechanics with manifest $\so(7)$ symmetry}
With our choice of the fermions $\psi_0,\psi_i$, the basic Poisson brackets between  the generators of $\mathcal{N}=8$ Poincar\'{e} supersymmetry read
\be\label{PB0}
\{Q_0,Q_0\}= 2 \im\, H, \; \{Q_0,Q_i\} =0, \qquad \{Q_i,Q_i\}= 2 \im\,\delta_{ij} H ,
\ee
To extend this algebra to the to superconformal one, we need the generators of dilatation ${\cal D}$, conformal boost ${\cal K}$ and superconformal transformations $S_i, S_0$ defined now as follows:
\be\label{conf2} 
{\cal D}= \frac{1}{2} r\,p_r, \; {\cal K}= \frac{1}{2} r^2, \quad S_{i} = r \psi_{i} , \; S_{0} = r \psi_{0} ,
\ee
where the new bosonic field $r$ and its momentum $p_r$ obey the standard brackets
\be
\{ p_r, r\} =1.
\ee
The basic Poisson brackets have the form
\be\label{ss}
\{S_{i},S_{j}\} = 2\,\im\, \delta_{ij} K , \quad 
\{S_{0},S_{0}\} = 2\,\im\, K , \qquad \{S_{i},S_{0}\} = 0.
\ee
Depending on the structure of superconformal symmetry, the generators of $R$-symmetry  include the bosonic  $\{V_i, W_{ij}\}$ and/or fermionic  $\so(8)$ generators $\{\hV_i, \hW_{ij}\}$ or some subset of these generators.

In full correspondence with our Ansatz for the      $\cN=8$ supercharges\footnote{Compare with  those  for the $\cN=4$ case \cite{tigran,SQS24}.} \cite{hkn1,KN1},  the supercharges must have the form :
\be
Q_{i} =p_r\,\psi_{i} + \frac{\Theta_{i}}{r},\qquad \qquad  Q_{0} = 
p_r\,\psi_{0}+\frac{ \Theta_{0}}{r} ,
\label{sc0}\nonumber
\ee
The composites 
$\Theta^{i\, A}$ and  $\Upsilon^{a\, \alpha}$ have the structure "$(R-symmetry\; generators) \times \, fermions$". Thus, the terms cubic in  fermions in the supercharges correspond to fermionic realizations of the $R$-symmetry generators while the terms linear in fermions in the supercharges correspond to the bosonic $R$-symmetry generators.

\subsection{$OSp(8|2)$ superconformal mechanics} 
The supercharges generating  the $OSp(8|2)$ superconformal mechanics read 
\bea\label{OSp}
&& Q_0 =  p_r \psi_0 -\frac{1}{4\, r} \left( c_{ijk} W_{ij} \psi_k + 2 V_i \psi_i \right), \nn\\
&& Q_i = p_r \psi_i + \frac{1}{4\,r} \left( 2 V_i \psi_0 -2 W_{ij}\psi_j+ 2 c_{ijk}V_j\psi_k + c_{ijk} W_{jk} \psi_0 - f_{ijkl}W_{jk} \psi_l \right).
\eea
The supercharges form the $\cN=8$ super Poincar\'{e} algebra
\be
\{Q_0,Q_0\}= 2 \im\, H, \; \{Q_0,Q_i\} =0, \qquad \{Q_i,Q_i\}= 2 \im\,\delta_{ij} H ,
\ee
with the Hamiltonian
\be\label{H1}
H=\frac{1}{2}p_r^2 +\frac{1}{r^2} \left[ \frac{1}{8} V_i(V_i -4 \hV_i) + \frac{1}{16} W_{ij}(W_{ij}-4 \hW_{ij}) -\frac{1}{4} c_{ijk}( V_i \hW_{jk} + \hV_i W_{jk})-
\frac{1}{8} f_{ijkl} W_{ij}\hW_{kl}\right]
\ee
The bosonic part of the Hamiltonian \p{H1} reads
\be
H_{bos}  = \frac{1}{2} p_r^2 +\frac{1}{16\,r^2} \sum_{\mu,\nu=0}^7 W_{\mu\nu}W_{\mu\nu} .
\label{osp82}\ee
It describes the particle on the eight-dimensional Euclidean cone, since the angular part of this Hamiltonian is the seven-dimensional sphere \cite{KN1}.

To understand the full dynamical symmetry of the system, one has to calculate the brackets between the Poincar\'e \p{OSp} and conformal supersymmetry generators \p{conf2}:
\bea
&& \{Q_0, S_0\}= 2 \im\, D, \nn \\ 
&& \{ Q_0, S_i\} =-\im \left( \hV_i+\frac{1}{2}V_i+\frac{1}{4} c_{ijk} W_{jk}\right) \equiv -\im\,\mathbf{V}_i, \quad \{ Q_i, S_0\} =\im\, \mathbf{V}_i, \nn \\
&& \{ Q_i, S_j\} =2 \im\, \delta_{ij}D -\im \left(\frac{1}{2} W_{ij}+\frac{1}{2}c_{ijk} V_k+\frac{1}{4} f_{ijkl}W_{kl}+\hW_{ij}\right) \equiv =2 \im\, \delta_{ij}D -\im\, \mathbf{W}_{ij}.
\eea
Thus we see that we have  $\so(8)$ $R$-symmetry spanned by the generators:
\bea 
\mathbf{V}_i  =  \hV_i+\frac{1}{2} \left( V_i +\frac{1}{2} c_{ijk} W_{jk}\right), \;
\mathbf{W}_{ij} = \hW_{ij} +\frac{1}{2}\left( W_{ij} +c_{ijk}V_k+\frac{1}{2} f_{ijkl}W_{kl}\right), \\
\{ \mathbf{W}_{ij},\mathbf{W}_{kl}\}  = \delta_{ik}\mathbf{W}_{jl}-\delta_{jk} \mathbf{W}_{il}-\delta_{il}\mathbf{W}_{jk}+\delta_{jl} \mathbf{W}_{ik},\;
\{ \mathbf{W}_{ij},\mathbf{V}_{k}\}  = \delta_{ik}\mathbf{V}_{j}-\delta_{jk} \mathbf{V}_{i},\;\{ \mathbf{V}_{i},\mathbf{V}_{j}\} =\mathbf{W}_{ij}.
\eea
The generators  $\mathbf{V}_i, \mathbf{W}_{ij}$ are conserved and they have the following Poisson brackets with the Hamiltonian \p{H1} and supercharges \p{OSp}
\bea
\{H,\mathbf{V}_i\} & = & \{H,\mathbf{W}_{ij}\} =0,\qquad \{\mathbf{V}_i,Q_0\} =Q_i, \;  \{\mathbf{V}_i,Q_j\} =-\delta_{ij} Q_0,\nn \\
\{\mathbf{W}_{ij},Q_k\} & = & \delta_{ik}Q_j-\delta_{jk}Q_i, \quad
\{\mathbf{W}_{ij},Q_0\} =0 .
\eea

\subsection{$\mathfrak{F}(4)$ superconformal mechanics} 
The supercharges generating  the $\mathfrak{F}(4)$ superconformal mechanics can be constructed with the help of $\so(7)$ generators $W_{ij}$ and $\hW_{ij}$ \p{so7} as follows
\bea\label{F4}
&& Q_0 =  p_r \psi_0 -\frac{1}{3\, r} c_{ijk}\left( W_{ij}  -\frac{1}{6}\hW_{ij} \right)\psi_k, \nn\\
&& Q_i = p_r \psi_i - \frac{1}{3\,r} \left( 2 W_{ij}\psi_j  -
c_{ijk}\left( W_{jk}-\frac{1}{2} \hW_{jk}\right) \psi_0 + f_{ijkl}\left( W_{jk}+\frac{1}{6}\hW_{jk}\right) \psi_l \right).
\eea
The supercharges form the $\cN=8$ super Poincar\'{e} algebra
\be
\{Q_0,Q_0\}= 2 \im\, H, \; \{Q_0,Q_i\} =0, \qquad \{Q_i,Q_i\}= 2 \im\,\delta_{ij} H ,
\ee
with the Hamiltonian
\be\label{H2}
H=\frac{1}{2}p_r^2 +\frac{1}{r^2} \left[ \frac{1}{9} W_{ij}\left(W_{ij} -3\hW_{ij}\right) - \frac{1}{6}f_{ijkl}\left(W_{ij}+\frac{1}{12} \hW_{ij}\right) \hW_{kl}-
\frac{1}{3} c_{ijk}\left( W_{ij} -\frac{1}{6} \hW_{ij}\right)\hV_j \right].
\ee
The bosonic part of the Hamiltonian \p{H2} reads
\be
H_{bos}  = \frac{1}{2} p_r^2 + \frac{1}{9\, r^2} \sum_{i,j=1}^7 W_{ij}W_{ij} ,
\label{317}\ee
In contrast with \eqref{osp82}, it  describes the particle on the seven-dimensional Euclidean cone, since the angular part of this Hamiltonian is the six-dimensional sphere \cite{KN1}.

To understand the full dynamical symmetry of the system, one has to calculate the brackets between the Poincar\'e \p{F4} and conformal supersymmetry generators \p{conf2}:
\bea
&& \{Q_0, S_0\}= 2 \im\, D, \nn \\ 
&& \{ Q_0, S_i\} =-\frac{\im}{3} c_{ijk}\widehat{\mathbf{W}}_{jk}, \quad
 \{ Q_i, S_0\} =\frac{\im}{3} c_{ijk}\widehat{\mathbf{W}}_{jk}, \nn \\
&& \{ Q_i, S_j\} =2 \im\, \delta_{ij}D -\frac{\im}{3} f_{ijkl}\widehat{\mathbf{W}}_{kl} -\frac{2\,\im}{3}\widehat{\mathbf{W}}_{ij},
\eea
where
\be
\widehat{\mathbf{W}}_{ij} = W_{ij}+\frac{1}{2}\left( \hW_{ij} +  c_{ijk}\hV_k+ \frac{1}{2} f_{ijkl}\hW_{kl} \right).
\ee
Thus we see that we have  $\so(7)$ $R$-symmetry spanned by the generators $\mathbf{W}_{ij}$. These generators obey the following brackets:
\be 
\{H,\widehat{\mathbf{W}}_{ij}\}  = 0,\qquad
\{\widehat{\mathbf{W}}_{ij},Q_0\}  =  \frac{1}{2} c_{ijk}Q_k , \quad
\{\widehat{\mathbf{W}}_{ij},Q_k\} =\frac{1}{2}\left( \delta_{ik}Q_j-\delta_{jk}Q_i\right) -\frac{1}{2}c_{ijk}Q_0 +
\frac{1}{2} f_{ijkl} Q_l .
\ee

\subsection{$G(3)$ superconformal mechanics with $\cN=7$ supersymmetry} 
To construct $G(3)$ superconformal mechanics, one has to explore the fermionic
version of the $\mathfrak{g}_2$ generators \p{G2}
\be
{\widehat\cW}_{ij} = \hW_{ij} +\frac{1}{4} f_{ijkl} \hW_{kl} .
\ee
With their help the unique solution for supercharges spanning $N=7$ super-Poincar\'{e} algebra
reads
\be \label{G3}
Q_i = p_r\, \psi_i - \frac{1}{6\, r} f_{ijkl} {\widehat\cW}_{jk} \psi_l = 
 p_r\, \psi_i -\frac{\im}{12\, r} f_{ijkl} \psi_{j} \psi_{k} \psi_{l}.
\ee
The supercharges \p{G3} form the $\cN=7$ super Poincar\'{e} algebra
\be
\{Q_i,Q_i\}= 2 \im\,\delta_{ij} H ,
\ee
with the Hamiltonian
\be\label{H3}
H= \frac{1}{2} p_r^2 - \frac{1}{18 r^2}  {\widehat\cW}_{ij}  {\widehat\cW}_{ij} = 
 \frac{1}{2} p^2+ \frac{1}{48 x^2} f_{ijkl} \psi_i \psi_j \psi_k \psi_l .
\ee

Again, to clarify the full dynamical symmetry of the system, one has to calculate the brackets between the Poincar\'e \p{G3} and conformal supersymmetry generators $S_i$ \p{conf2}:
\be
\{ Q_i, S_j\} =2 \im\, \delta_{ij} D - \im\, {\widehat\cW}_{ij}.
\ee
Thus we see that we have  $\mathfrak{g}_2$ $R$-symmetry spanned by the generators ${\widehat\cW}_{ij}$. These generators obey the following brackets:
\bea
&&\{H,{\widehat\cW}_{ij}\} =0, \quad \{{\widehat\cW}_{ij},Q_k\} =  \delta_{ik}Q_j -\delta_{jk}Q_i +
\frac{1}{2} f_{ijkl}Q_l , \\
&& \{{\widehat\cW}_{ij},S_k\} =  \delta_{ik} S_j -\delta_{jk}S_i +
\frac{1}{2} f_{ijkl}S_l  , \nn \\
&& \{{\widehat\cW}_{ij}, {\widehat\cW}_{kl}\} =-\frac{3}{2} \left( f_{ijkm} {\widehat\cW}_{lm} - f_{ijlm}{\widehat\cW}_{km} \right) +
 c_{ijm}c_{kln} {\widehat\cW}_{mn} .
\eea

Alas, we did not find any possibility to include in the game more bosonic fields. So, for the time being, we have $\cN=7$ superconformal mechanics with $G(3)$ superconformal dynamical symmetry with one bosonic and seven fermionic components.

\setcounter{equation}0
\section{Superconformal mechanics with manifest $\su(2)$ symmetry}
With our choice of the fermions \p{conj}, the basic Poisson brackets between  the generators of $\mathcal{N}=8$ Poincar\'{e} supersymmetry read
\be\label{N8P}
\left\{Q^{ i\,A} , Q^{j\, B} \right\} = 4 \imath\epsilon^{ij} \epsilon^{AB} {\cal H}, \quad   \left\{q^{a\,\alpha }, q^{b\, \beta}\right\} = 4   \imath \epsilon^{ab } \epsilon^{\alpha \beta} \, {\cal H}, \qquad
\left\{Q^{i\,A} , q^{a\,\alpha} \right\} = 0. 
\ee
To extend this algebra to the  superconformal ones, we need the generators of dilatation ${\cal D}$, conformal boost ${\cal K}$, and superconformal transformations $S^{i\,A}, s^{a\,\alpha}$ defined as follows:
\be\label{conf3} 
{\cal D}= \frac{1}{2} r\,p_r, \; {\cal K}= \frac{1}{2} r^2, \quad S^{i\,A} = r \phi^{i\,A} , \; 
s^{a\,\alpha} = r \chi^{a\, \alpha} .
\ee
The basic Poisson brackets have the form\cite{hkn1,KN1}
\be\label{N8Conf}
\left\{S^{ i\,A} , S^{j\, B} \right\} = 4 \imath\epsilon^{ij} \epsilon^{AB} {\cal K}, \quad   \left\{s^{a\,\alpha }, s^{b\, \beta}\right\} = 4   \imath \epsilon^{a b } \epsilon^{\alpha \beta} \, {\cal K}, \qquad
\left\{S^{i\,A} , s^{a\,\alpha} \right\} = 0. 
\ee
Depending on the structure of superconformal symmetry, the generators of $R$-symmetry  include the bosonic  $\{J^{ij}, W^{AB}, X^{ab}, Y^{\alpha\beta}, V^{i\,A\,a\,\alpha}\}$ \p{su2bos},  and/or fermionic  $\so(8)$ generators $\{\hJ^{ij}, \hW^{AB}, \hX^{ab}, \hY^{\alpha\beta}, \hV^{i\,A\,a\,\alpha}\}$ \p{su2fer}, \p{so8} or some subset of these generators.

In full correspondence with our Ansatz for the $\cN=8$ supercharges \cite{hkn1,KN1} (see  \cite{tigran,SQS24} for the $\cN=4$ case), the supercharges have the form :
$$
Q^{i\,A} =p_r\phi^{i\,A} + \frac{\Theta^{i\,A}}{r},\qquad \qquad  q^{a\, \alpha} = 
p_r\chi^{a\, \alpha}+\frac{ \Upsilon^{a\, \alpha}}{r} .
%\label{sc}
$$
The composites 
$\Theta^{i\, A}$ and  $\Upsilon^{a\, \alpha}$ have the structure "$(R-symmetry\; generators) \times \, fermions$". 
%Thus, the terms cubic in  fermions in supercharges correspond to fermionic realizations of %$R$-symmetry generators while terms linear in fermions in supercharges correspond to %bosonic $R$-symmetry generators.

\subsection{$OSp(8|2)$ superconformal mechanics} 
This superconformal mechanics was constructed in \cite{fed1} at the level of superfields and components. The corresponding supercharges and  Hamiltonian for this case were constructed in \cite{hkn1,KN1}. Their have the form
\bea\label{scosp}
Q^{i\, A} & =&  p_r \phi^{i\, A} +\frac{1}{r}\left( J^{ij}\phi_j{}^A+  W^{AB}\phi^i{}_B - \frac{1}{2} V^{i\,A\, a\,\alpha} \chi_{a\, \alpha}, \right), \nn \\
q^{a\,\alpha}& = & p_r \chi^{a\, \alpha}+\frac{1}{r}\left( X^{ab}\chi_b{}^\alpha+  Y^{\alpha\beta} \chi^a{}_\beta+ \frac{1}{2} V^{i\,A\, a\,\alpha} \phi_{i\, A}\right).
\eea

The corresponding Hamiltonian reads:
\be\label{hso8}
H_{Osp(8|2)} = \frac{1}{2} p_r^2 +\frac{1}{r^2} \left( -\frac{1}{2} {\cal C}_{\so(8)} +\frac{5}{8} ({\cal C}_{\so(8)}|_{fermions \rightarrow 0})\right) .
\ee 
The $\so(8)$ Casimir operator ${\cal C}_{\so(8)}$ is defined in a standard way as
\bea 
{\cal C}_{\so(8)}& \equiv &\left( J^{ij}+  \hJ^{ij}\right)\left( J_{ij}+  \hJ_{ij}\right) +\left( W^{AB}+  \hW^{AB}\right)\left( W_{AB}+  \hW_{AB}\right) + \left( X^{ab}+  \hX^{abj}\right)\left( X_{ab}+  \hX_{ab}\right)+ \nn \\
&& + \left( Y^{\alpha\beta}+ \hY^{\alpha\beta}\right)\left( Y_{\alpha\beta}+\hY_{\alpha\beta}\right)+\frac{1}{4} \left( V^{i\, A\, a\, \alpha }+\hV^{i\, A\, a\, \alpha }\right)\left( V_{i\, A\, a\, \alpha} +\hV_{i\, A\, a\, \alpha}\right) .
\eea

The generators of  dynamical $\so(8)$ $R$-symmetry of the system appear in the brackets between
the Poincar\'e \p{scosp} and conformal supersymmetry generators \p{conf3}:
\bea\label{spconf}
\left\{Q^{i\,A}, S^{j\, B}\right\} & = & 2\, \im\, \epsilon^{ij} \left( W^{AB}+
\hW^{AB}\right) + 2\, \im\, \epsilon^{AB} \left( J^{ij} + \hJ^{ij}\right) + 4\, \im\, \epsilon^{ij}\epsilon^{AB} D, \nn\\
\left\{Q^{i\,A}, s^{a\, \alpha}\right\} & = & -\, \im\, \left( V^{i\,A\,a\,\alpha}+\hV^{i\,A\,a\,\alpha}\right), \quad  
\left\{q^{a\, \alpha}, S^{i\, A}\right\}  =  \im\, \left( V^{i\,A\, a\, \alpha}+
\hV^{i\,A\, a\, \alpha}\right),   \nn \\
\left\{q^{a\,\alpha}, s^{b\, \beta}\right\} & = & 2\, \im\, \epsilon^{ab} \left(Y^{\alpha\beta}+\hY^{\alpha\beta}\right) + 2 \, \im \, \epsilon^{\alpha\beta}\left( X^{ab}+\hX^{ab}\right) +
4\, \im\,\epsilon^{ij}\epsilon^{\alpha \beta} D .
\eea

The supercharges $Q^{i\,A}$ and $q^{a\,\alpha}$ form the fundamental representations with respect to the corresponding $\su(2)$ algebras, i.e.
\bea
&& \{J^{ij}+\hJ^{ij},Q^{k\,A}\} =-\frac{1}{2}\left(\epsilon^{ik} Q^{j\,A}+\epsilon^{jk} Q^{i\,A}\right), \;
\{W^{AB}+\hW^{AB},Q^{i\,C}\} = -\frac{1}{2}\left(\epsilon^{AC}Q^{i\,B}+\epsilon^{BC}Q^{i\,A} \right),\nn\\
&& \{X^{ab}+\hX^{ab},Q^{i\,A}\} = \{Y^{\alpha\beta}+\hY^{\alpha\beta},Q^{i\,A}\} = 0,\;  \{J^{ij}+\hJ^{ij},q^{a\,\alpha}\} =\{W^{AB}+\hW^{AB},q^{a\,\alpha}\}=0,   \nn\\
&& \{X^{ab}+\hX^{ab},q^{c\,\alpha}\} =-\frac{1}{2}\left(\epsilon^{ac} q^{b\,\alpha}+\epsilon^{bc} q^{a\,\alpha}\right),\; \{Y^{\alpha\beta}+\hY^{\alpha\beta},q^{a\,\gamma}\} =-\frac{1}{2}\left(\epsilon^{\alpha\gamma} q^{a\,\beta}+\epsilon^{\beta\gamma} q^{a\,\alpha}\right).
\eea
In addition, these currents commute with the Hamiltonian \p{hso8}:
\be
\{H_{Osp(8|2)},J^{ij}+\hJ^{ij} \} =\{H_{Osp(8|2)},W^{AB}+\hW^{AB}\} =\{H_{Osp(8|2)},X^{ab}+\hX^{ab} \} =\{H_{Osp(8|2)},Y^{\alpha\beta}+\hY^{\alpha\beta}\} =0.
\ee
Finally, we have
\bea
&&\{V^{i\,A\,a\,\alpha}+\hV^{i\,A\,a\,\alpha},Q^{j\,B}\} = 2 \epsilon^{ij}\epsilon^{AB} q^{a\,\alpha},\;\{V^{i\,A\,a\,\alpha}+\hV^{i\,A\,a\,\alpha},q^{b\,\beta}\} = -2 \epsilon^{ab}\epsilon^{\alpha\beta} Q^{i\,A},\\ 
&& \{H_{Osp(8|2)},V^{i\,A\,a\,\alpha}+\hV^{i\,A\,a\,\alpha}\} =0. \nn
\eea

The bosonic part of the Hamiltonian \p{hso8}  then coincides with   \eqref{osp82}  \footnote{The $\so(8)$ operators $\ell_{\mu,\nu}$
	are defined in \p{ell}. Note, the term $ \sum_{\mu,\nu=1}^8 \ell_{\mu,\nu} \ell_{\mu,\nu} $ does not fix the norm  $y_\mu y_\mu$, which therefore can be identified with $r^2$ reducing "evident" nine bosonic coordinates to eight.} 
\be\label{ospbos}
H_{bos} =  \frac{1}{2} p_r^2 +\frac{1}{8 r^2}  {\cal C}_{\so(8)}|_{fermions \rightarrow 0} =
\frac{1}{2} p^2 +\frac{1}{16 r^2} \sum_{\mu,\nu=1}^8 \ell_{\mu,\nu} \ell_{\mu,\nu}  .
\ee 

\subsection{${\mathfrak F}(4)$ supersymmetric mechanics}
Now to construct the supercharges we have to use the generators  spanning $\so(7)$ algebra, i.e.
$$J^{ij}, Y^{\alpha\beta}, Z^{ab}_{\so(7)}, V^{i\,a\,b\,\alpha}_{\so(7)} $$
with $ Z^{ab}_{\so(7)}$ and $V^{i\,a\,b\,\alpha}_{\so(7)}$ defined in \p{zso7},\p{vso7}. In addition, the fermions $\phi^{i\,A}$ carrying out now the
indices $\{i,a\}$, i.e. $\phi^{i\,A} \rightarrow \phi^{i\,a}$.
    
The {${\mathfrak F}(4)$ supersymmetric mechanics is described by the following supercharges:
\bea\label{F4}
&& Q^{i\,a} =  p_r \phi^{i\,a} -\frac{1}{r} \left[ \left(\frac{2}{3} Z^{a\,b}_{\so(7)}+2 \hZ^{a\,b}_{\so(7)} \right) \phi^i{}_b +\frac{4}{9}\left(  3 J^{ij} +5 \hJ^{ij}\right) \phi_j{}^a + \frac{1}{3}\left( V_{\so(7)}^{i\,a\,b\,\alpha}+\hV_{\so(7)}^{i\,a\,b\,\alpha}\right)\chi_{a \alpha}\right], \nn \\
&& q^{a\,\alpha} =  p_r \chi^{a\,\alpha} +\frac{1}{r} \left[\left(\frac{2}{3} Z^{a\,b}_{\so(7)}+2 \hZ^{a\,b}_{\so(7)}\right) \chi_b{}^\alpha 
+\frac{4}{9} \left(3 Y^{\alpha \beta} +5 \hY^{\alpha\,\beta}\right) \chi^a_\beta
-  \frac{1}{3}\left( V_{\so(7)}^{j\,a\,b\,\alpha}+\hV_{\so(7)}^{j\,a\,b\,\alpha}\right)\phi_{j b}\right],
\eea
and by the Hamiltonian
\be\label{HF4}
H_{{\mathfrak F}(4)} =\frac{1}{2} p_r^2+ \frac{1}{r^2}\left[ -\frac{2}{3}{\cal C}_{so(7)}+\frac{8}{9} \left({\cal C}_{so(7)}|_{fermions\rightarrow 0}\right)+\frac{4}{9} \left({\cal C}_{so(7)}|_{bosons\rightarrow 0}\right) \right]
.
\ee
Here ${\cal C}_{\so(7)}$ is the Casimir operator commuting with the generators $\{ J^{ij}+\hJ^{ij}, Z^{ab}_{\so(7)}+\hZ^{ab}_{\so(7)},Y^{\alpha\beta}+\hY^{\alpha\beta}, V^{i\,a\,b\,\alpha}_{\so(7)}-\hV^{i\,a\,b\,\alpha}_{\so(7)}\}$:
\bea\label{casso7}
{\cal C}_{so(7)} &=&\left(J^{ij}+\hJ^{ij}\right)\left( J_{ij}+\hJ_{ij}\right)+\frac{1}{2} \left(Z^{ab}_{\so(7)} +\hZ^{ab}_{\so(7)} \right)\left((Z_{\so(7)})_{ab}+(\hZ_{\so(7)})_{ab}\right)+ \left(Y^{\alpha\beta }+
\hY^{\alpha\beta}\right)\left( Y_{\alpha\beta}+ \hY_{\alpha\beta}\right) + \nn\\
&& \frac{1}{16}\left(  V^{i\,a\,b\,\alpha}_{\so(7)}- \hV^{i\,a\,b\,\alpha}_{\so(7)}\right) \left( (V_{\so(7)})_{i\,a\,b\,\alpha}-
 (\hV_{\so(7)})_{i\,a\,b\,\alpha}\right).
\eea

To understand the full dynamical symmetry of the system, one has to calculate the Poisson brackets between
the Poincar\'e \p{F4} and conformal supersymmetry generators \p{conf3} (with the indices $A,B$ replaced by the indices $a,b$):
\bea\label{F4conf}
\left\{Q^{i\,a}, S^{j\, b}\right\} & = & \frac{4}{3}\, \im\, \epsilon^{ij}\left( Z^{ab}_{\so(7)}+\hZ^{ab}_{\so(7)}\right) + \frac{8}{3}\, \im\, \epsilon^{ab} \left(J^{ij}+\hJ^{ij}\right)+4\, \im\, \epsilon^{ij}\epsilon^{ab} D, \nn\\
\left\{Q^{i\,a}, s^{b\,\alpha}\right\} & = & \frac{2}{3}\, \im\, \left( V^{i\,a\,b\,\alpha}_{\so(7)} - \hV^{i\,a\,b\,\alpha}_{\so(7)}\right), \quad  
\left\{q^{b\,\alpha}, S^{i\,a}\right\}  = -\frac{2}{3}\, \im\, \left( V^{i\,a\,b\,\alpha}_{\so(7)} - \hV^{i\,a\,b\,\alpha}_{\so(7)}\right),   \nn \\
\left\{q^{a\,\alpha}, s^{b\,\beta}\right\} & = & \frac{4}{3}\, \im\, \epsilon^{\alpha\beta} \left( Z^{a\,b}_{\so(7)}+\hZ^{a\,b}_{\so(7)}\right)+ \frac{8}{3} \, \im \, \epsilon^{ab} \left( Y^{\alpha\beta}+ \hY^{\alpha\beta}\right) +
4\, \im\,\epsilon^{ij}\epsilon^{ab} D .
\eea

Thus, we see that instead of $\so(8)$ symmetry we indeed have  its subalgebra spanned by the  generators
\be\label{so7b}
\{ J^{ij}+\hJ^{ij}, Z^{ab}_{\so(7)}+\hZ^{ab}_{\so(7)}, Y^{\alpha\beta}+\hY^{\alpha\beta},  V^{i\,a\,b\,\alpha}_{\so(7)}- \hV^{i\,a\,b\,\alpha}_{\so(7)} \}.
\ee
These 21 generators span $\so(7)$ subalgebra in $\so(8)$. 

Thus, we deal with ${\mathfrak F}(4)$ superconformal symmetry. Note, the pure fermionic parts of the supercharges in \p{F4} coincide with those obtained in \cite{F4scm}. 

The generators \p{so7b} are the conserved quantities 
\be
\{H_{{\mathfrak F}(4)}, \{J^{ij}+\hJ^{ij}, Z^{ab}_{\so(7)}+\hZ^{ab}_{\so(7)}, Y^{\alpha\beta}+\hY^{\alpha\beta},  V^{i\,a\,b\,\alpha}_{\so(7)}- \hV^{i\,a\,b\,\alpha}_{\so(7)}\}\} =0
\ee
and they rotate the supercharges as follows:
\bea
&& \{J^{ij}+\hJ^{ij},Q^{k,\,a}\} =-\frac{1}{2} \left(\epsilon^{ik} Q^{j\,a}+
\epsilon^{jk} Q^{i\,a} \right), \;  \{J^{ij}+\hJ^{ij},q^{a\,\alpha}\} =0, \nn \\
&& \{Z^{ab}_{\so(7)}+\hZ^{ab}_{\so(7)},Q^{i\,c}\} = -\frac{1}{2} \left( \epsilon^{ac}Q^{i\,b}+\epsilon^{bc}Q^{i\,a} \right),  \;
 \{Z^{ab}_{\so(7)}+\hZ^{ab}_{\so(7)},q^{c\,\alpha}\} = -\frac{1}{2} \left( \epsilon^{ac}q^{b\,\alpha}+\epsilon^{bc}q^{a\,\alpha} \right), \nn \\
&&  \{Y^{\alpha\beta}+\hY^{\alpha\beta },Q^{i\,a}\} =0, \;
 \{Y^{\alpha\beta}+\hY^{\alpha\beta },q^{a\,\gamma}\} = -\frac{1}{2}\left( \epsilon^{\alpha\gamma} q^{a\, \beta}+\epsilon^{\beta\gamma} q^{a\, \alpha}\right), \\
&& \{V^{i\,a\,b\,\alpha}_{\so(7)}- \hV^{i\,a\,b\,\alpha}_{\so(7)}, Q^{j\,c} \} =
 -2 \epsilon^{ij}\left(\epsilon^{ac} q^{b\,\alpha}+\epsilon^{bc} q^{a\,\alpha}\right), \;
 \{V^{i\,a\,b\,\alpha}_{\so(7)}- \hV^{i\,a\,b\,\alpha}_{\so(7)}, q^{c\,\beta} \} =
 2 \epsilon^{\alpha\beta}\left(\epsilon^{ac} Q^{i\,b}+\epsilon^{bc} Q^{i\, a}\right). \nn
\eea

The bosonic part of the Hamiltonian \p{HF4} then coincides with \eqref{317} 
\be\label{f4bos}
H_{bos} =  \frac{1}{2} p_r^2 +\frac{2}{9 r^2}  {\cal C}_{so(7)}|_{fermions \rightarrow 0} =
\frac{1}{2} p^2_r +\frac{1}{9 r^2} \sum_{\mu,\nu=1}^7 \ell_{\mu,\nu} \ell_{\mu,\nu}  .
\ee 

\subsection{$G(3)$ superconformal algebra}
Now to construct the supercharges spanning the $G(3)$ superconformal algebra,  we have to use the generators spanning ${\mathfrak g}_2$ algebra, i.e.
$$J^{ij},  Z^{ab}_{\g_2}, V^{i\,a\,b\,ca}_{\g_2} $$
with $ Z^{ab}_{\g_2}$ and $V^{i\,a\,b\,c}_{\g_2}$ defined in \p{G21},\p{G22}. In addition, the fermions $\phi^{i\,A}$ carrying  now the
indices $\{i,a\}$, i.e. $\phi^{i\,A} \rightarrow \phi^{i\,a}$, and the fermions 
$\chi^{a\,\alpha}$ carrying the indices  $\{a,b\}$ are symmetric over these indices, i.e. 
\be
\chi^{i\,A} \rightarrow \chi_{\g_2}^{a\,b} =\chi^{a\,b}+\chi^{b\,a}.
\ee
Correspondingly, we have the supercharges $Q^{i\,a}, q^{a\,b} = q^{b\,a}$ with the following Poisson brackets:
\be\label{G3q}
\{Q^{i\,a },Q^{j\,b}\}  =  4\, \im \,\epsilon^{ij} \epsilon^{ab} H , \quad
\{q^{ab},q^{cd}\}  =  8\,\im\, \left( \epsilon^{ac} \epsilon^{bd} +\epsilon^{ad}\epsilon^{cb}\right) H ,\quad \{Q^{i\,a},q^{b\,c}\} =0.
\ee

The conformal supercharges \p{conf2} can also be re-defined as
\bea
&& S^{i\,a} = r \, \phi^{i\,a}, \qquad s^{a\,b} = r\, \chi^{ab}_{\G_2}, \label{conf4} \\
&& \{S^{i\,a },S^{j\,b}\}  =  4\, \im \,\epsilon^{ij} \epsilon^{ab} K , \quad
\{s^{ab},s^{cd}\}  =  8\,\im\, \left( \epsilon^{ac} \epsilon^{bd} +\epsilon^{ad}\epsilon^{cb}\right) K ,\quad \{S^{i\,a},s^{b\,c}\} =0.
\eea
The supercharges spanning $\cN=7$ super Poincare algebra have the form
\bea\label{Qq}
Q^{i\,a} & = & p_r\, \phi^{i\,a} +\frac{1}{r} \left[ \frac{1}{3} \hJ^{i\,j} \phi_j{}^a - \frac{1}{32}
\hV^{i\,a\,b\,c}_{\g_2}(\chi_{\g_2})_{b\,c} \right], \nn \\
q^{a\,b} & = & p_r\,\chi^{a\,b}_{\g_2}+\frac{1}{8 r} \hV^{j\,c\,a\,b}_{\g_2}\phi_{j\,c} 
.
\eea
while the Hamiltonian $H$ reads
\be\label{G3H}
H_{\G(3)}=\frac{1}{2} p^2_r-\frac{1}{r^2} \left[\frac{5}{12} \hJ^{ij}\hJ_{ij} +\frac{1}{4} \hZ^{ab}_{\g_2} \left(\hZ_{\g_2}\right){}_{ab}\right] = \frac{1}{2} p^2_r+\frac{1}{192\,r^2}\left[3 \phi^{ia}\phi_i^b (\chi_{\g_2})_a{}^c  (\chi_{\g_2})_{b c}+
2 \phi^{ia}\phi_i{}^b \phi^j{}_a \phi_{j b}\right]
\ee

To understand the full dynamical symmetry of the system, one has to calculate the Poisson brackets between the Poincar\'e \p{Qq} and conformal supersymmetry generators \p{conf4}:
\bea
&& \{Q^{i\,a},S^{j\,b}\}=- \im\, \epsilon^{ij} \hZ^{ab}_{\g_2} - 3 \im \epsilon^{ab} \hJ^{ij} +4 \im\, \epsilon^{ij} \epsilon^{ab} D,\nn \\
&& \{q^{ab},S^{i\,c}\} = \frac{1}{2} \im\, \hV^{i\,a b c}_{\g_2},\;
\{Q^{i\, a},s^{b\,c}\} = - \frac{1}{2} \im\, \hV^{i\,a b c}_{\g_2}, \nn \\
&& \{q^{ab}, s^{cd}\} = 4\, \im \left(\epsilon^{ac} \hZ^{bd}_{\g_2} + \epsilon^{bd} \hZ^{ac}_{\g_2}\right) +
8\, \im \,\left( \epsilon^{ac}\epsilon^{bd}+\epsilon^{ad}\epsilon^{bc}\right) D.
\eea
Thus we see that we have $R$-symmetry spanned by $\su(2) \oplus \su(2)$ algebras with the generators $\hZ^{ab}_{\g_2}=\hX^{ab}+ \hW^{ab} +\hY^{ab}$ and $\hJ^{ij}$ and the generators $\hV^{i a b c}_{\g_2}$ from the coset
$\g_2/\su(2)\times \su(2)$. Therefore, we are dealing with $G(3)$ superconformal symmetry.

The generators  $\hJ^{ij},\hZ^{ab}_{\g_2}$ are the conserved quantities, i.e. they commute with the Hamiltonian \p{G3H}:
\be
\{H_{\G(3)},\hJ^{ij}\} = \{H_{\G(3)},\hZ^{ab}_{\g_2}\}=0.
\ee
The Poisson brackets between the generators  $\hJ^{ij},\hZ^{ab}_{\g_2}$ and the supercharges \p{Qq} read
\bea
&& \{\hJ^{ij},Q^{k\,a}\} = -\frac{1}{2}\left( \epsilon^{ik} Q^{j\,a}+ \epsilon^{jk} Q^{i\,a}
\right),\;  \{\hJ^{ij},q^{a\,b}\} = 0, \nn \\
&& \{\hZ_{\g_2}^{ab}, Q^{i\,c} \} =-\frac{1}{2} \left( \epsilon^{ac}Q^{i\,b}+\epsilon^{bc}Q^{i\,a}\right), \; \{\hZ_{\g_2}^{ab}, q^{c\,d} \}=
-\frac{1}{2} \left(\epsilon^{ac} q^{b\,d}+\epsilon^{ad} q^{b\,c}+\epsilon^{bc} q^{a\,d}+\epsilon^{bd} q^{a\,c}\right), \nn \\
&& \{ \hV^{i\,a\,b\,c}_{\g_2},Q^{j\,d}\} =2 \epsilon^{ij}\left( \epsilon^{ad} q^{b\,c}+\epsilon^{bd} q^{a\,c}+\epsilon^{cd} q^{a\,b}\right), \nn \\
&& \{ \hV^{i\,a\,b\,c}_{\g_2},q^{d\,e}\} = -4\left[ \epsilon^{ad} \left( \epsilon^{be} Q^{i\,c}+
\epsilon^{ce} Q^{i\,b} \right)+
\epsilon^{bd} \left( \epsilon^{ae} Q^{i\,c}+ \epsilon^{ce} Q^{i\,a} \right)+
\epsilon^{cd} \left( \epsilon^{ae} Q^{i\,b}+
\epsilon^{be} Q^{i\,a} \right) \right].
\eea
\section{Conclusion}
In the present paper we have constructed the supercharges and Hamiltonians for two variants of  superconformal mechanics associated with the superalgebras $OSp(8|2), \mF(4)$ and
$G(3)$. The fermionic and bosonic fields involved were arranged into the generators spanning $\so(8), \so(7)$ and $\g_2$ $R$-symmetries of the  corresponding superconformal algebras. The bosonic and fermionic parts of these  $R$-symmetry generators   define the constants of motion and form the same  algebras. We have also presented the explicit embedding of
$\g_2$ into $\so(7)$ for both realizations, with manifest $\so(7)$ and $\su(2)$ symmetries. These embeddings have never been presented in such form to the best of our knowledge.

Similarly to the results of our previous papers \cite{hkn1,KN1,SQS24}, the   angular parts of supercharges defining the system have the structure ``$(R-symmetry\;generators)\, \times\, fermions$'', while the angular part of the Hamiltonian is just a sum of  full  Casimir operators and its purely bosonic and/or fermionic parts. We explicitly demonstrated that the constructed system describes the specific $\cN=7,8$ supersymmetric extension of  free particles on some cone  embedded in  (pseudo-)Euclidean space with the 	fermionic part that can be interpreted as the spin-orbit coupling terms. The $\cN=7$ superconformal mechanics with dynamical $G(3)$ symmetry describe the first example of the classical system
constructed on the non-minimal supermultiplet $(1,7,7,1)$. It is very interesting to construct a superspace description of this system. 
All our mechanics admit the reduction: $bosonic\quad R \quad  symmetry\;  generators \; \rightarrow 0$. Probably, these cases are the simplest ones for the search of superfield
description.

Finally, we would like to stress that the explicit form of the supercharges and Hamiltonians
contains the generators of the bosonic parts of $R$-symmetry. The structure of these bosonic
generators is more or less arbitrary. Thus, similarly to the $\cN=4$ supersymmetric case
\cite{anton} we have the possibility to describe a different bosonic system simply by choosing  different realizations of these symmetries. Clearly, our supercharges in this case will provide the $\cN=8$ super-extension for these systems. We are planning to consider this situation in more detail in forthcoming publications.

\def\theequation{A.\arabic{equation}}

\section*{Acknowledgements} 
The authors  thank Alexander Molev, Oleg Ogievetsky, Francesco Toppan and Alexei Morozov  for useful comments.  

The work of A.N.  was partially supported by the Armenian State Committee of Higher Education and Science, project    21AG-1C062.

The work of S.K.  was partially supported by the Ministry of Science and Higher Education of Russia, Government Order for 2023-2025, Project No. FEWM-2023-0015 (TUSUR).

\setcounter{equation}0
\section*{Appendix}
To treat the bosonic variables  $\{B^{a A}{}_m, Z^{i \alpha}{}_m\}$  as additional coordinates and momenta in the first order Lagrangians, one has  to introduce these new coordinates and momenta in the supersymmetric mechanics we are dealt with. The simplest way is to associate the fields $\{B^{a A}{}_2, Z^{i \alpha}{}_2\}$ with lower index $m=2$ with the momenta and the fields with lower index $m=1$, i.e.  $\{B^{a A}{}_1, Z^{i \alpha}{}_1\}$, with the coordinates. One of the possible associations read
\bea
&& B^{11}{}_1=y^1 + \im y^2, \;B^{12}{}_1=y^3 + \im y^4, \;B^{21}{}_1= -y^3 +\im y^4, \; B^{22}{}_1=y^1 - \im y^2,  \nn \\
&& B^{11}{}_2=-p_1 - \im p_2, \;B^{12}{}_2=-p_3-\im p_4, \; B^{21}{}_2 =p_3 -\im p_4, \; B^{22}{}_2 = -p_1+\im p_2, \label{BBb} \\
&& Z^{11}{}_1= y^5+\im y^6, \; Z^{12}{}_1 = y^8+\im y^7, \; \; Z^{21}{}_1=-y^8+\im y^7, \; Z^{22}{}_1=y^5-\im y^6, \nn \\
&& Z^{11}{}_2=-p_5-\im p_6, \; Z^{12}{}_2=-p_8-\im p_7, \; Z^{21}{}_2= p_8-\im p_7, \; Z^{22}{}_2=-p_5+\im p_6 .\label{zz}
\eea
One can check that with the standard brackets
\be
\{p_\mu,y_\nu\} =\delta_{\mu\nu} ,\qquad \mu,\nu=1,\ldots, 8
\ee
the brackets \p{PB} are correctly reproduced.

In terms of the coordinates $y^\mu$ and momenta $p_\mu$, the $\so(8)$ generators have the standard form
\be\label{ell}
\ell_{\mu,\nu} = p_\mu y_\nu - p_\nu y_\mu .
\ee

\end{document}